\documentstyle[12pt]{article}

\newlength{\dinwidth}
\newlength{\dinmargin}
\def\be{\begin{equation}}
\def\ee{\end{equation}}
\def\bea{\begin{eqnarray}}
\def\eea{\end{eqnarray}}

\begin{document}
\begin{titlepage}
\begin{flushright}
CERN-TH-7337/94\\
FTUAM-94/15\\
June, 1994\\
hep-th/9406206
\end{flushright}
\bigskip
\begin{center}
{\LARGE A Canonical Approach to Duality Transformations}
\vskip 0.9truecm

{Enrique \'Alvarez
\footnote{On leave of absence from: Departamento de F\'{\i}sica
Te\'orica,
Universidad Aut\'onoma de Madrid, 28049 Madrid, Spain},
Luis \'Alvarez-Gaum\'e
and  Yolanda Lozano \footnotemark[1]}

\vspace{1pc}

{\em Theory Division CERN \\1211 Geneva 23\\ Switzerland}\\

\vspace{5pc}

{\large \bf Abstract}
\end{center}

We show that Buscher's abelian duality transformation rules
can be recovered in a very simple way by performing a canonical
transformation first suggested by Giveon, Rabinovici and Veneziano.
We explore the properties of this transformation, and also
discuss some aspects of non-abelian duality.

\vfill

\end{titlepage}

\def\theequation{\thesection . \arabic{equation}}

\section{Introduction}
\setcounter{equation}{0}
(Target space) duality, the generalization of the well-known
$R$-duality in toroidal compactifications in String Theory
\cite{kikawa},
remains a promising avenue
to further
understand the underlying symmetries of String Theory.

The most powerful description of this symmetry was first
introduced by
Buscher \cite{buscher}, and then further elaborated upon by
many others \cite{mogollon} \cite{reviewkiritsis}
\cite{aagbl} \cite{quevedo} \cite{aagl}.
It starts with the sigma-model
formulation of the
corresponding conformal field theory (``string vacuum''),
and it works when the target space metric has at
least one continuous
isometry. Even in this case, however, the procedure looks
unnecessarily complicated: the isometry is gauged, the
(non propagating) gauge fields are constrained to be trivial,
and the Lagrange multipliers themselves are promoted to the
rank of new coordinates once the gaussian integration over the
gauge fields is performed. One suspects that all those
complicated intermediate steps could be avoided, and that it
should be possible to pass directly from the original to the
dual theory.

Some suggestions have indeed been made in the literature
pointing (at least in the simplified situation where
all backgrounds are constant or dependent only on t)
towards an understanding
of duality as particular instances of canonical
transformations \cite{venezia}.

Our aim in this paper is to follow this lead.
We shall find that it works well when the isometry is
abelian,
laying the theory on a simpler
setting than before, namely as a (privileged) subgroup of
the whole group of (non-anomalous, that is implementable
in Quantum Field Theory \cite{ghandour}, \cite{mascanonicas}
\cite{masmas})
canonical transformations on the phase space of the theory.

We shall be able to perform the transformation
starting from arbitrary coordinates (and writing the dual manifold
in arbitrary coordinates as well) and, besides, we will also
follow explicitly the transformations of the currents of both
the initial and the dual model.  In particular in the case
of WZW models \cite{WZW} it becomes rather simple to prove
that the full duality group is given by $Aut(G)_L\times
Aut(G)_R$, where $L,R$ refer to the left- and right-currents
on the model with group $G$, and $Aut(G)$ are the automorphisms
of G, both inner and outer.  Due to the chiral conservation
of the currents in this case, the canonical transformation
leads to a local expression for the dual currents.  In the
case where the currents are not chirally conserved, then
those currents associated to symmetries not commuting with
the one used to perform duality become generically non-local
in the dual theory and this is why they are not manifest
in the dual Lagrangian \cite{zachos}.

It is, however, somewhat disappointing to have to report that we
have not been able to understand non-abelian duality in terms
of canonical transformations. Once again in the case of
WZW-models, due to the presence of chirally conserved currents
one can compute the dual theory in terms of a set
of non-local variables introduced in \cite{aagl}. This procedure
unfortunately only works for theories with chiral currents,
and even for that case we lack a Hamiltonian formulation
of the transformation.

\section{Abelian Duality and Canonical Transformations}
\setcounter{equation}{0}

We start with the bosonic
sigma model written in arbitrary coordinates
on a manifold $M$ with Lagrangian
\be
\label{sigmamodel}
L = \frac12 (g_{\mu\nu}+b_{\mu\nu})(\phi) \partial_{+}\phi^\mu
\partial_{-}\phi^\nu
\ee
where $x^\pm=(\tau\pm\sigma)/2$, $\mu,\nu=1,\dots ,n={\rm dim}M$.
The corresponding Hamiltonian is
\be
H=\frac12 (g^{\mu\nu} (p_\mu-b_{\mu\rho}\phi^{'}\,^\rho)
(p_\nu-b_{\nu\sigma}\phi^{'}\,^\sigma) + g_{\mu\nu}\phi^{'}\,^\mu
\phi^{'}\,^\nu)
\ee
where $\phi^{'}\,^\mu\equiv d\phi^\mu/d\sigma$.
We assume, moreover, that there is a Killing vector field $k^\mu$,
${\cal L}_k g_{\mu\nu}=0$ and $i_k H=-dv$ for some one-form $v$,
where $(i_k H)_{\mu\nu}\equiv k^\rho H_{\rho\mu\nu}$
and $H=db$ locally.
This guarantees the existence of a particular
system of coordinates (``adapted coordinates''), which we shall
denote by $x^a\equiv (\theta,x^i)$, such that
$\vec{k}=\partial / \partial\theta$. We shall denote the
jacobian matrix by $e^a_\mu\equiv\partial x^a/
\partial \phi^{\mu}$. In practice, this means that we have to
complete the Killing vector $\vec{k}\equiv k^\mu \partial /
\partial \phi^\mu$ with $(n-1)$ other commuting vectors in order
to form a holonomic n-bein: $[\vec{k},\vec{e_i}]=[\vec{e_i},
\vec{e_j}]=0$.

This then defines a point transformation
in the original Lagrangian (\ref{sigmamodel})
which acts on the Hamiltonian as a canonical transformation
with generating
function $\Phi=x^a(\phi)p_a$, which yields:
\bea
p_\mu=e^a_\mu p_a \nonumber\\
x^a=x^a(\phi)
\eea
Once in adapted coordinates we can write the sigma model
Lagrangian as
\be
L=\frac12 G (\dot{\theta}^2-\theta^{'}\,^2)+(\dot{\theta}
+\theta^{'})J_-+
(\dot{\theta}-\theta^{'})J_++V
\ee
where
\bea
G=g_{00}=k^2 \qquad V=\frac12 (g_{ij}+b_{ij})\partial_+
x^i\partial_-x^j
\nonumber\\
J_-=\frac12 (g_{0i}+b_{0i})\partial_-x^i \qquad J_+=\frac12
(g_{0i}-b_{0i})\partial_+x^i
\eea
In finding the dual with a canonical
transformation we can use the Routh function with respect
to $\theta$,
i.e. we only apply the Legendre transformation to
$(\theta,\dot{\theta})$. The canonical momentum is given by
\be
p_\theta=G\dot{\theta}+(J_++J_-)
\ee
and the Hamiltonian
\bea
&&H=p_\theta \dot{\theta}-L=\frac12 G^{-1} p_\theta^2-
G^{-1}(J_++J_-)p_\theta+\frac12 G\theta^{'}\,^2+ \nonumber\\
&&+\frac12 G^{-1}(J_++J_-)^2+\theta^{'}(J_+-J_-)-V.
\eea

The Hamilton equations are:
\bea
\dot{\theta}=\frac{\delta H}{\delta p_\theta}=
G^{-1}(p_\theta-J_+-J_-) \nonumber\\
\dot{p_\theta}=-\frac{\delta H}{\delta\theta}=
(G\theta^{'}+J_+-J_-)^{'}
\eea
where the last equation is equivalent to current conservation.

The current components are:
\bea
{\cal J_+}=\frac12 G \partial_+\theta+J_+=\frac12 p_\theta+\frac12
G\theta^{'}+\frac{J_+-J_-}{2} \nonumber\\
{\cal J_-}=\frac12 G \partial_-\theta+J_-=\frac12 p_\theta-\frac12
G\theta^{'}-\frac{J_+-J_-}{2}
\eea

Then $\partial_-{\cal J_+}+\partial_+{\cal J_-}=0 \Leftrightarrow
\dot{p_\theta}=-\delta H/ \delta\theta$. The generator of the
canonical transformation we choose is:

\be
\label{fungen}
F=\frac12 \int_{D, \partial D=S^1} d\tilde{\theta}\wedge d\theta=
\frac12 \oint_{S^1}
(\theta^{'}\tilde{\theta}-\theta\tilde{\theta}^{'}) d\sigma
\ee
that is,
\bea
p_\theta=\frac{\delta F}{\delta\theta}=-\tilde{\theta}^{'} \nonumber\\
p_{\tilde{\theta}}=-\frac{\delta F}{\delta\tilde{\theta}}=-\theta^{'}
\eea
This generating functional does not receive any quantum corrections
(as explained in \cite{ghandour}) since it is linear in $\theta$ and
$\tilde{\theta}$.  It $\theta$ were not an adapted coordinate to
a continuous isometry, the canonical transformation would
generically lead to a non-local form of the dual Hamiltonian.
Since the Lagrangian and Hamiltonian in our case only depend
on the time- and space-derivatives of $\theta$, there are no
problems with non-locality.  Below we will show how to
transform wave functionals from the original to the dual theory.

The dual Hamiltonian is:
\bea
\tilde{H}=\frac12 G^{-1}
\tilde{\theta}^{'}\,^2+G^{-1}(J_++J_-)\tilde{\theta}^{'}+ \nonumber\\
\frac12 G p_{\tilde{\theta}}^2-(J_+-J_-)p_{\tilde{\theta}}+
\frac12 G^{-1} (J_++J_-)^2-V
\eea

Since:
\be
\dot{\tilde{\theta}}=\frac{\delta\tilde{H}}{\delta
p_{\tilde{\theta}}}=Gp_{\tilde{\theta}}-(J_+-J_-),
\ee
we can perform the inverse Legendre transform:
\bea
&&\tilde{L}=\frac12 G^{-1} (\dot{\tilde{\theta}}^2-
\tilde{\theta}^{'}\,^2)+
G^{-1}J_+(\dot{\tilde{\theta}}-\tilde{\theta}^{'}) \nonumber\\
&&-G^{-1}J_-(\dot{\tilde{\theta}}+\tilde{\theta}^{'})+
V-2G^{-1}J_+J_-.
\eea
When translated in terms of $g_{ab},b_{ab}$ this is nothing but
Buscher's transformations\footnote{The minus
signs in $\tilde{g}_{0i}$ and $\tilde{b}_{0i}$ can be absorbed
in a redefinition $\tilde{\theta}\rightarrow -\tilde{\theta}$.}
\bea
\label{buscher}
&&\tilde{g}_{00}=1/g_{00},\qquad
         \tilde{g}_{0i}=-b_{0i}/g_{00},\qquad
          \tilde{g}_{ij} = g_{ij} -
\frac{g_{0i}g_{0j} - b_{0i} b_{0j}}{g_{00}}\nonumber\\
&&\tilde{b}_{0i} = -\frac{g_{0i}}{g_{00}},\qquad
        \tilde{b}_{ij}=b_{ij}-\frac{g_{0i}b_{0j}
         -g_{0j}b_{0i}}{g_{00}}
\eea
For the dual theory to be conformal invariant the dilaton
must transform as $\Phi^{'}=\Phi-\log{g_{00}}$
\cite{buscher} \cite{ema}.

\vspace*{1cm}

Some useful information can be extracted easily in the
approach of the canonical transformation:

-From the generating functional (\ref{fungen}) we can learn
about the multivaluedness and periods of the dual variables
\cite{aagbl}.
Since $\theta$ is periodic and in the path integral
the canonical transformation is implemented by \cite{ghandour}:
\be
\psi_k[\tilde{\theta}(\sigma)]=N(k)\int {\cal D}\theta(\sigma)
e^{iF[\tilde{\theta},\theta(\sigma)]}
\phi_k[\theta(\sigma)]
\ee
where $N(k)$ is a normalization factor,
$\phi_k(\theta+a)=\phi_k(\theta)$ implies for $\tilde{\theta}$:
$\tilde{\theta}(\sigma+2\pi)-\tilde{\theta}(\sigma)=4\pi /a$,
which means that
$\tilde{\theta}$ must live in the dual lattice of $\theta$.  Note
that (2.16) suffices to construct the dual Hamiltonian.
It is a simple exercise to check that acting with (2.12)
on the left-hand side of (2.16) and pushing the dual
Hamiltonian through the integral we obtain the original
Hamiltonian acting on $\phi_k[\theta(\sigma)]$:
\be
\tilde{H}\psi_k[\tilde{\theta}(\sigma)]=N(k)\int {\cal D}
\theta(\sigma) e^{iF[\tilde{\theta},\theta(\sigma)]}
H\phi_k[\theta(\sigma)]
\ee
This makes the duality
transformation very simple conceptually,
and it also implies how it can be applied to arbitrary
genus Riemann surfaces, because the state
$\phi_k[\theta(\sigma)]$ could be the state obtained
by integrating the original theory on an arbitrary
Riemann surface with boundary.  It is also clear that
the arguments generalize straightforwardly when we
have several commuting isometries.

-We can easily see that under the canonical transformation the
Hamilton equations are interchanged:
\bea
\dot{p_\theta}=-\frac{\delta H}{\delta\theta}=
(G\theta^{'}+J_+-J_-)^{'} \rightarrow
\dot{\tilde{\theta}}=Gp_{\tilde{\theta}}-J_++J_- \nonumber\\
\dot{\theta}=\frac{\delta H}{\delta p_\theta}=
G^{-1}(p_\theta-J_+-J_-) \rightarrow
\dot{p_{\tilde{\theta}}}=(G^{-1}(\tilde{\theta}^{'}+J_++J_-))^{'}
\eea
The canonical transformed current conservation law is
in this case equivalent to the first Hamilton equation:
\be
\partial_-{\cal J_+}^{c.t.}+\partial_+{\cal J_-}^{c.t.}=0
\Leftrightarrow \dot{\tilde{\theta}}=\frac{\delta\tilde{H}}{\delta
p_{\tilde{\theta}}}
\ee
(where we denote the canonical transformed current by
${\cal J}_\pm^{c.t.}$)
whereas the dual current\footnote{Note that the canonical transformed
current does not coincide in general with the dual current.}
conservation law is equivalent to the
second Hamilton equation.

-In the chiral case $J_-=0$ (i.e. $g_{0i}=-b_{0i}$) and $G$ is
a constant, therefore we can
normalize $\theta$ to set
$G=1$ and :
\be
L=\frac12  (\dot{\theta}^2-\theta^{'}\,^2)+
(\dot{\theta}-\theta^{'})J_++V
\ee
The Hamiltonian is
\be
H=\frac12 p_\theta^2-J_+p_\theta+\frac12 (J_++\theta^{'})^2-V
\ee

The action is invariant under $\delta\theta=\alpha(x^+)$,
a $U(1)_L$ Kac-Moody symmetry.
The $U(1)$ Kac-Moody algebra has the
automorphism ${\cal J}_+\rightarrow -{\cal J}_+$. This is
precisely the effect of the canonical transformation.
The equation of motion or current conservation is:
\be
\partial_-(\partial_+\theta+J_+)=0
\ee
${\cal J}_+=\partial_+\theta+J_+=p_\theta+\theta^{'}$ transforms
under the canonical transformation in
${\cal J}_+^{c.t.}=-\tilde{\theta}^{'}-
p_{\tilde{\theta}}=-{\cal J}_+$.

-We can also follow the transformation to the dual model
of other continuous symmetries.  The simplest case is
as usual the WZW-model \cite{WZW} which is the basic
model with chiral currents.  Consider for simplicity
the level-$k$ $SU(2)$-WZW model with action
\be
S[g]={-k\over 2\pi}\int d^2 \sigma
Tr (g^{-1}\partial_+g g^{-1}\partial_-g)
+{k\over 12\pi}\int Tr(g^{-1}dg)^3,
\ee
parametrizing $g$ in terms of Euler angles,
\be
g=e^{i\alpha\sigma_3/2}e^{i\beta\sigma_1/2}e^{i\gamma\sigma_3/2}
\ee
the left- and right-chiral currents are
\be
{\cal J}_+={k\over 2\pi}\partial_+g g^{-1}\qquad
{\cal J}_-=-{k\over 2\pi}g^{-1}\partial_-g.
\ee
Explicitly for ${\cal J}_+$:
\bea
\label{corrientes}
&&{\cal J}^1_+={k\over 2\pi}(-\cos\alpha\sin\beta
\partial_+\gamma+\sin\alpha\partial_+\beta)\nonumber\\
&&{\cal J}^2_+={k\over 2\pi}(\sin\alpha\sin\beta
\partial_+\gamma+\cos\alpha\partial_+\beta)\nonumber\\
&&{\cal J}^3_+={k\over 2\pi}(\partial_+\alpha +
\cos\beta\partial_+\gamma),
\eea
and similarly for the right currents.  If we perform
duality with respect to $\alpha\rightarrow\alpha +{\rm constant}$
we know that ${\cal J}^3_+\rightarrow- {\cal J}^3_+,
{\cal J}^3_-\rightarrow {\cal J}^3_-$.  For these currents it is
easy to find the action of the canonical transformation
because only the derivatives of $\alpha$ appear.  For
${\cal J}^{1,2}_+$ there is an explicit dependence on $\alpha$
and it seems that the transform of these currents is very
non-local.  However due to the chiral nature of these
currents, we can show now that there are similar chirally
conserved currents in the dual model.  To do this we
first combine the currents in terms of root generators,
\bea
\label{root}
&&{\cal J}^{(+)}_+={\cal J}^1_+ +i{\cal J}^2_+=e^{-i\alpha}
(i\partial_+\beta-\sin\beta\partial_+\gamma)
=e^{-i\alpha}j^{(+)}_+, \nonumber\\
&&{\cal J}^{(-)}_+={\cal J}^1_+ -i{\cal J}^2_+=-e^{i\alpha}
(i\partial_+\beta+\sin\beta\partial_+\gamma)=
e^{i\alpha}j^{(-)}_+,
\eea
from chiral current conservation $\partial_-{\cal J}^{(\pm)}_+
=0$ we obtain
\be
\partial_- j^{(\pm)}_+=\pm i\partial_-\alpha j^{(\pm)}_+,
\ee
in these equations only $\dot\alpha,\alpha'$ appear, and
after the canonical transformation we can reconstruct the
dual non-abelian currents ( in the previous equations
the canonical transformation amounts to the replacement
$\alpha\rightarrow\tilde{\alpha}$) which take the same
form as the original ones except that with respect to
the transformed ${\cal J}^3_+$ the roles of positive and
negative roots get exchanged.  One also verifies that
${\cal J}^a_-$ are unaffected.  This implies therefore
that the effect of duality with respect to shifts
of $\alpha$ is an automorphism of the current algebra
amounting to performing a Weyl transformation on the
left currents only while the right ones remain unmodified.
This result although known \cite{reviewkiritsis} is much
easier to derive in the Hamiltonian formalism
than in the Lagrangian formalism where one must
introduce external sources which carry some ambiguities.
The construction for $SU(2)$ can be straightforwardly
extended to other groups.  This implies that for
WZW-models the full duality group is
$Aut(G)_L\times Aut(G)_R$, where $Aut(G)$ is the group
of automorphisms of the group $G$, including Weyl
transformations and outer automorphisms.  For instance
if we take $SU(N)$, the transformation
$J_+\rightarrow -J_+^{T}$, i.e. charge conjugation,
follows from a canonical transformation of the
type discussed.  It suffices to take as generating
functions for the canonical transformation the sum
of the generating functions for each generator
in the Cartan subalgebra.  External automorphisms
have been used recently to extend the notion of
duality \cite{bs} \cite{ao} \cite{oberskiritsis}.
It is important
to remark that the chiral conservation of the
currents is crucial to guarantee the locality
of the dual non-abelian currents.  If the conserved
current with respect to which we dualize is not
chirally conserved locality is
not obtained.  The simplest example to verify
this is the principal chiral model for $SU(2)$.
Although it is not a CFT, it serves for
illustrative purposes. The equations of motion for
this model imply the conservation laws:
\be
\partial_-{\cal J}^a_++\partial_+{\cal J}^a_-=0
\ee
where
\be
{\cal J}^a_\pm=\frac{k}{2\pi}\partial_\pm g g^{-1}
\ee
If we perform duality with respect to the invariance under
$\alpha$ translations we know how ${\cal J}^3_\pm$
transform, since they are the currents associated to the
isometry. With the canonical transformation is possible
to see as well which are the other dual conserved currents.
Since the dual model is only $U(1)$-invariant \cite{aagbl}
one expects the rest of the currents to become non-local
\cite{zachos}
if they exist at all.
In terms of the root generators introduced in (\ref{root})
the conservation laws
\be
\partial_-{\cal J}^{(\pm)}_++\partial_+{\cal J}^{(\pm)}_-=0
\ee
are expressed:
\be
\partial_-j^{(\pm)}_++\partial_+j^{(\pm)}_-
\mp i(\partial_-\alpha j^{(\pm)}_++
\partial_+\alpha j^{(\pm)}_-)=0
\ee
Performing the canonical transformation we obtain that
the dual conserved currents are given by:
\bea
{\tilde {\cal J}}^{(+)}_{\pm}=\exp{(i\int d\sigma
(\dot{\tilde{\alpha}}+\cos\beta \gamma^{'}))}
(i\partial_{\pm}\beta-\sin\beta\partial_{\pm}\gamma) \nonumber\\
{\tilde {\cal J}}^{(-)}_{\pm}=-\exp{(-i\int d\sigma
(\dot{\tilde{\alpha}}+\cos\beta \gamma^{'}))}
(i\partial_{\pm}\beta+\sin\beta\partial_{\pm}\gamma)
\eea
which cannot be expressed in a local form.

\vspace*{1.5cm}

The dual manifold $\tilde{M}(\tilde{\theta},x^i)$ is automatically
expressed in coordinates adapted to the dual Killing vector
$\tilde{\vec{k}}=\partial / \partial\tilde{\theta}$. We can now
perform
another point transformation, with the same jacobian as before
\be
\tilde{e}^a_\mu=e^a_\mu
\ee
to express the dual manifold in coordinates which are as close as
possible to the original ones.

The transformations we have performed are: First a point
transformation $\phi^\mu\rightarrow \{\theta, x^i\}$, to go
to adapted coordinates in the original manifold. Then a
canonical transformation $\{\theta, x^i\}\rightarrow
\{\tilde{\theta}, x^i\}$,
which is the true duality transformation. And finally another
point transformation $\{\tilde{\theta}, x^i\}\rightarrow
\tilde{\phi}^\mu$,
with the same jacobian as the first point transformation,
to express
the dual manifold in general coordinates.

It turns out that the composition of these three transformations
can be expressed in geometrical terms using only the Killing
vector
$k^\mu$, $\omega_\mu\equiv e^0_\mu$ and the corresponding dual
quantities \footnote{Please note that we must raise and
lower indices
with the dual metric, i.e. $\tilde{e}_{a\mu}=\tilde{g}_{ab}
\tilde{e}^{b}_\mu; \tilde{e}^{a\mu}=\tilde{g}^{\mu\nu}
\tilde{e}^{a}_\nu$,
and this implies $\tilde{\omega}_{\mu}=\omega_{\mu}$, but
$\tilde{\omega}^{\mu}=k^{\mu} (k^2+v^2)+\vec{e}\,^{\mu}
\cdot v$ (where
$\vec{e}\,^{\mu}\equiv e^{\mu}_i$);
$\tilde{k}^{\mu}=k^{\mu}$, but
$\tilde{k}_{\mu}=(\omega_{\mu} -
(\vec{e}_{\mu}\cdot v))/k^2$; we have
moreover $\tilde{\omega}\,^2=
k^2+v^2+g^{ij} v_{j}{\omega}_{i}$ and
$\tilde{k}^2=1/k^2$.}.

It is then quite easy to check that the total canonical
transformation
to be made in (\ref{sigmamodel}) is just

\bea
&&k^\mu p_\mu \rightarrow \tilde{\omega}_\mu
\tilde{\phi}^{'}\,^\mu
\nonumber\\
&&\omega_\mu \phi^{'}\,^\mu \rightarrow \tilde{k}^\mu
\tilde{p}_\mu
\eea
whose generating function is\footnote{The one-form
$\omega\equiv \omega_\mu d\phi^\mu$ is dual to
the Killing vector $\vec{k}$: $\omega(\vec{k})=1$,
$\omega(\vec{e}_i)=0$; but it is of course
different from $\underline{k}\equiv k_\mu /k^2 d\phi^\mu$
(the former is an exact form, whereas the latter does
not even in general satisfy Frobenius condition
$\underline{k}\wedge d\underline{k}=0$).}
\be
F=\frac12 \int_D \tilde{\omega}\wedge\omega=
\frac12 \int_D\tilde{\omega}_\mu d\tilde{\phi}^\mu\wedge
\omega_{\rho}d\phi^{\rho}
\ee

One then easily performs the transformations in such a way that
the dual metric and torsion can be expressed in geometrical terms
as\footnote{In checking that the total result is equivalent to
Buscher's transformations, one needs to use the fact that
$\omega^2=\frac{1}{k^2}+g^{ij}k_ik_j /k^4$.}
\be
\tilde{g}_{\mu\nu}=g_{\mu\nu}-\frac{1}{k^2}(k_\mu k_\nu-
(v_\mu -\omega_\mu)(v_\nu-\omega_\nu))
\ee
\be
\tilde{g}^{\mu\nu}=g^{\mu\nu}+\frac{1}{(1+k.v)^2}
[(k^2+(v-\omega)^2)k^{\mu} k^{\nu} -2(1+k.v)(k^{(\mu}
(v-\omega)^{\nu)}]
\ee
and
\be
\tilde{b}_{\mu\nu}=b_{\mu\nu}-\frac{2}{k^2}k_{[\mu}
(v-\omega)_{\nu]}
\ee
which are the generalization of Buscher's formulae (\ref{buscher})
in arbitrary
coordinates, where
\bea
k_{(\mu} (v-\omega)_{\nu)}=\frac12 (k_\mu(v_\nu-\omega_\nu)+
k_\nu(v_\mu-\omega_\mu))\nonumber\\
k_{[\mu} (v-\omega)_{\nu]}=\frac12 (k_\mu(v_\nu-\omega_\nu)-
k_\nu(v_\mu-\omega_\mu))
\eea

\section{Non-abelian Duality}
\setcounter{equation}{0}

In view of the simplicity of the canonical approach to abelian
duality, one could be tempted to think that the corresponding
generalization to the non-abelian case would not be very
difficult. Unfortunately this is not the case, the reason
being that there are no adapted coordinates to
a set of non-commuting isometries, and therefore
one is led to a non-local form of the Hamiltonian.
In \cite{aagl} we could carry out the non-abelian
duality transformation due to the existence of left-
and right-chiral currents and as a consequence of
the Polyakov-Wiegmann \cite{polywieg} property satisfied
by WZW-actions.  Although in the intermediate steps
it was necessary to introduce non-local variables
(a convenient representation of the auxiliary
variables setting the non-abelian gauge field
strength to zero), the final result led to a
local action in the new variables as a result
of the special properties of WZW-models mentioned.
The computations could be carried out exactly until
the end to evaluate the form of the effective
action in terms of the auxiliary variables needed
in the construction of non-abelian duals.  Although a
glance to the action could lead us to think that
the non-abelian dual of a WZW-model with
group $G$  with respect to a non-anomalous subgroup
$H$ is locally a coset theory $G/H$ times a WZW-theory
for the subgroup $H$, a careful analysis of the
BRST constraints that appear in the construction
together with a computation of the duality transformation
for toroidal world-sheets show that the model is
self-dual.  We have so far been unable to express
these functional integral manipulations in a Hamiltonian
setting as in previous section.  We should mention
however that the use of non-local variables
introduced in \cite{aagl} in cases where the currents
are conserved but not chiraly conserved leads
in general to non-local actions, hence they are not
very useful in those cases.  A setting where they are
still useful is in the case of $(1,1)$-supersymmetric
WZW-models \cite{vkpr} \cite{tseytlin} \cite{maharana}.
We will not
spell out the details here because the results
are exactly the same as in the bosonic case
\cite{aagl} except that fields are replaced
by $(1,1)$-superfields, and that the shift
of level $k\rightarrow k+c_H$ ($c_H$ is the dual
Coxeter number of the subgroup $H$) does not take
place due to cancellations between bosonic
and fermionic determinants.  Apart from these cases
a satisfactory treatment of non-abelian duality
in non-chiral cases in the functional integral
or Hamiltonian approaches still remains elusive.

\section{Conclusions}

We have presented in this paper a canonical approach to (target
space) duality in String Theory. We believe that this is a
``minimal'' approach in the sense that no extraneous structure
needs to be introduced, and all standard results in the abelian
case (and more) are easily recovered using it. In particular,
the behavior of currents not commuting with those used
to implement duality is clarified.  All the generators
of the full duality group $O(d,d,Z)$ can be described
in terms of canonical transformations.
This gives the impression that the duality group
should be understood in terms of global
symplectic diffeomorphisms.  The configuration space
of the two-dimensional field theories considered
is the loop space of a manifold with isometries,
the phase space is the cotangent space to this loop
space. It would be useful to formulate the notion
of duality in the context of some analogue of the
group of disconnected diffeomorphisms, but for the
time being we have not found such a construction.
Finally non-abelian duality seems to
fall beyond the scope of the Hamiltonian point
of view advocated in this paper.  Whether this
is just a question of technical ingenuity
or whether it is due to some conceptual difficulty
in the very notion of non-abelian duality remains
for the time being undecided.

\bigskip

{\large\bf Acknowledgements}

We would like to thank E. Kiritsis and M.A. V\'azquez-Mozo
for useful discussions.
E.A. and Y.L. were supported in part
by the CICYT grant AEN 93/673 (Spain) and by a fellowship of
Comunidad Aut\'onoma de Madrid (YL).

\bigskip

{\large\bf Note Added}

After completion of this work we learned that in reference
\cite{zachos}
a canonical transformation was built relating the $SU(2)$
principal
chiral model and its non-abelian dual with respect to the
left action
of the whole group. We thank T. Curtright and C. Zachos for
bringing this
information to our attention and for other discussions.
We also learned that this
non-abelian dual model was studied in
\cite{fj}.


\newpage

\begin{thebibliography}{99}

\medskip

\bibitem{kikawa}
K. Kikkawa and M. Yamasaki, Phys. Lett. B149
(1984) 357;\quad
N. Sakai and I. Senda, Prog. Theor. Phys. 75 (1984) 692;\quad
L. Brink,
M.B. Green and J.H. Schwarz, Nucl. Phys. B198
 (1982) 474.

\bibitem{buscher}
T.H. Buscher, Phys. Lett. B194 (1987) 51, B201 (1988) 466.

\bibitem{ema}
E. \'Alvarez and M.A.R. Osorio, Phys. Rev. D40 (1989)
1150.

\bibitem{mogollon}
M. Ro\u{c}ek and E. Verlinde, Nucl. Phys. B373
(1992) 630;
\quad
A. Giveon and M. Ro\u{c}ek, Nucl. Phys.
B380 (1992) 128;
\quad
A. Giveon and E. Kiritsis, Nucl. Phys. B411 (1994) 487;\quad
A. Giveon, M. Porrati and E. Rabinovici, {\it Target
Space Duality in String Theory},
NYU-TH-94/01/01.

\bibitem{reviewkiritsis}
E. Kiritsis, Nucl. Phys. B405 (1993) 109.

\bibitem{aagbl}
E. \'Alvarez, L. \'Alvarez-Gaum\'e, J.L.F.
Barb\'on and Y. Lozano, Nucl. Phys. B415 (1994) 71.

\bibitem{quevedo}
X. de la Ossa and F. Quevedo, Nucl. Phys.
B403 (1993) 377;\quad
A. Giveon and M. Ro\u{c}ek, {\it On Nonabelian Duality},
ITP-SB-93-44, hep-th/9308154;\quad
M. Gasperini, R. Ricci and G. Veneziano,
Phys. Lett. B319 (1993) 438.

\bibitem{aagl}
E. Alvarez, L. Alvarez-Gaum\'e and Y. Lozano,
{\it On Non-abelian Duality}, CERN-TH 7204/94,
hep-th/9403155. To appear in Nucl. Phys. B.

\bibitem{venezia}
A. Giveon, E. Rabinovici and G. Veneziano, Nucl. Phys. B322
(1989) 167; \quad K.A. Meissner and G. Veneziano,
Phys. Lett. B267
(1991) 33.

\bibitem{ghandour}
G.I. Ghandour, Phys. Rev. D35 (1987) 1289;\quad
T. Curtright, {\it Differential Geometrical Methods in Theoretical
Physics: Physics and Geometry}, ed. L.L. Chau and W. Nahm,
Plenum, New York, (1990) 279;\quad
T. Curtright and G. Ghandour, {\it Using Functional Methods to
compute Quantum Effects in the Liouville Theory}, proceedings of
NATO Advanced Workshop, Coral Gables, FL, Jan 1991; and references
therein.

\bibitem{mascanonicas}
A. Anderson, Phys. Lett. B305 (1993) 67;\quad
A.Y. Shiekh, J. Math. Phys. 29 (1988) 913.

\bibitem{masmas}
Canonical transformations have also been used in String Theory in:
J. Maharana and G. Veneziano, Nucl. Phys. B283
(1987) 126;\quad J. Maharana and S. Mukherji, Phys. Lett. B284
(1992) 36.

\bibitem{WZW}
E. Witten, Comm. Math. Phys. 92 (1984) 455.

\bibitem{polywieg}
A.M. Polyakov, P.B. Wiegmann, Phys. Lett. B131 (1983) 121,
B141 (1984) 223.

\bibitem{vkpr}
P. Di Vecchia, V. Knizhnik, J. Peterson and P. Rossi,
Nucl. Phys. B253 (1985) 701.

\bibitem{tseytlin}
A. Giveon, E. Rabinovici and A.A. Tseytlin,
Nucl. Phys. B409 (1993) 339.

\bibitem{maharana}
A. Das and J. Maharana, {\it Duality of the Superstring
in Superspace}, UR-1338, hep-th/9401147.

\bibitem{bs}
I. Bars and K. Sfetsos, Mod. Phys. Lett. A7 (1992) 1091.

\bibitem{ao}
I. Antoniadis and N. Obers, {\it Plane Gravitational
Waves in String Theory}, CPTH-A299.0494, hep-th/9403191.

\bibitem{oberskiritsis}
E. Kiritsis and N. Obers, {\it A New Duality Symmetry in String
Theory}, CERN-TH 7310/94, hep-th/9406082.

\bibitem{zachos}
T. Curtright and C. Zachos, Phys. Rev. D49 (1994) 5408.

\bibitem{fj}
B.E. Fridling and A. Jevicki, Phys. Lett. B134 (1984) 70.

\end{thebibliography}
\end{document}